\documentclass[reprint,prb,aps]{revtex4-1} 

\usepackage{graphicx}
\usepackage{amsmath}
\usepackage{amssymb}
\usepackage{bbm}
\usepackage{braket}
\usepackage{hyperref}

\newcommand{\diff}{\mathrm{d}}
\newcommand{\der}[1]{\frac{\mathrm{d}}{\mathrm{d}#1}}

\newcommand{\acom}[2]{\left\{#1,#2\right\}}

\newcommand{\ptrace}[2]{\text{Tr}_{#1}\left[ #2 \right]}
\newcommand{\unitmatrix}{\mathbbm{1}}

\newcommand{\cbar}[1]{\,\overline{#1}}

\newcommand{\order}[1]{\mathcal{O}\left(#1\right)}
\newcommand{\ketbra}[1]{\ket{#1}\bra{#1}}
\newcommand{\Ebar}{\overline{\hspace{-0.15em}E}}

\newcommand{\figref}[1]{Fig.~\ref{#1}}

\newcommand{\equref}[1]{Eq.~(\ref{#1})}
\newcommand{\equsref}[2]{Eqs.~(\ref{#1}) and (\ref{#2})}

\newcommand{\equstref}[2]{Eqs.~(\ref{#1} - \ref{#2})}
\newcommand{\secref}[1]{Sec.~\ref{#1}}
\newcommand{\secsref}[2]{Secs.~\ref{#1} and \ref{#2}}
\newcommand{\appref}[1]{Appendix~\ref{#1}}
\newcommand{\refcite}[1]{Ref.~\onlinecite{#1}}
\newcommand{\refscite}[1]{Refs.~\onlinecite{#1}}
\newcommand{\ie}{i.e.~}
\newcommand{\eg}{e.g.~}

\renewcommand{\Re}[1]{\text{Re}\left[#1\right]}
\renewcommand{\Im}[1]{\text{Im}\left[#1\right]}
\renewcommand{\approx}{\simeq}

\begin{document}
\title{Non-adiabatic processes in Majorana qubit systems}
\author{M.~S.~Scheurer}
\author{A.~Shnirman}
\affiliation{Institute for Theory of Condensed Matter and DFG Center for Functional Nanostructures (CFN), Karlsruhe Institute of Technology (KIT), Karlsruhe, Germany}
\date{\today}

\begin{abstract}
We investigate the non-adiabatic processes occurring during the manipulations of 
Majorana qubits in 1-D semiconducting wires with proximity induced superconductivity. 
Majorana qubits are usually protected by the excitation gap. Yet, manipulations 
performed at a finite pace can introduce both decoherence and renormalization effects.  
Though exponentially small for slow manipulations, these effects are important as they 
may constitute the ultimate decoherence mechanism.  Moreover, as adiabatic topological 
manipulations fail to produce a universal set of quantum gates, non-adiabatic manipulations 
might be necessary to perform quantum computation.  
\end{abstract}

\maketitle
\section{Introduction}
Various realizations of zero energy Majorana bound states (MBS) are currently being 
intensively investigated.\cite{Alicea} Initially introduced in rather abstract models\cite{Kitaev,Ivanov} they started to look 
realistic after several heterostructures that might host such modes were proposed.\cite{FuKane1D,Oreg, Lutchyn}
In all these heterostructures superconductivity is proximity induced into a semiconductor with 
strong spin-orbit coupling\cite{Oreg, Lutchyn} or into a surface state of a topological insulator.\cite{FuKane1D}

Two MBS form a regular (Dirac) fermion which can be  either occupied or non-occupied. 
These states are of different fermion parity and, thus, cannot be used as a qubit. However, setups 
with more MBS, e.g,. two pairs of MBS, are already rich enough to encode qubits within the 
subspace of a given parity. Topological manipulations of these qubits require "braiding" of MBS. 
In the simplest realization one just "mechanically" moves one MBS around another. This can be achieved 
by applying time-dependent gates.\cite{vonOppen} More sophisticated braiding schemes have been suggested (see \eg \refscite{Flensberg,vanHeck}).

In this paper we study the non-adiabatic effects occurring when the MBS are "mechanically" shifted. 
Yet, the formalism introduced here, is rather general and can be applied in more involved setups. 
Decoherence effects in Majorana qubits have already been addressed.\cite{Trauzettel,Goldstein,Cheng}
Coupling to a gapless fermonic bath is definitely detrimental.\cite{Trauzettel}. In \refcite{Goldstein,Cheng}  
a general framework of decoherence in situations when the gap is preserved, e.g., adiabatic manipulations, 
was introduced. In contrast to \refcite{Goldstein,Cheng}, 
we perform the adiabatic perturbation expansion for a concrete physical system calculating the non-adiabatic coupling 
matrix (Berry matrix) explicitly. 

We distinguish two major effects. First, due to the motion of the MBS a quasiparticle in the continuum may be excited, 
somewhat analogously to the Landau-Zener tunneling. The probability of such an event is exponentially small, unless 
the velocity of the MBS approaches a certain critical velocity. Such a process changes the parity of the qubit subspace, 
thus giving rise to decoherence.  Second, the coupling between two remote Majorana modes can get renormalized if both MBS are moved simultaneously. 
This coupling lifts the degeneracy between the empty and the occupied 
states of the corresponding Dirac fermion. 
On the one hand, this renormalization effect has, thus, to be accounted for, if we aim at performing quantum gates with high accuracy.
On the other hand, it can be generated intentionally in order to induce non-topological phase gates. 

This paper is organized as follows: In \secref{General_expressions} we first present the formalism that is used in this work for treating non-adiabatic effects, investigate a general time-dependent topological superconductor of class $D$ and derive an effective Hamiltonian for the qubit incorporating non-adiabatic corrections. Then we focus on the quantum wire proposal and discuss both the limitations arising from the presence of the states above the gap of the system (see \secref{Limitations}) as well as the possibility of creating a phase gate that is based on non-adiabatic effects (see \secref{UTQC}).

\section{General expressions for non-adiabatic processes}
\label{General_expressions}
In this section non-adiabatic processes in systems hosting MBS are analyzed from a generic point of view, \ie without referring to any of the specific realizations of Majorana modes in condensed matter systems.

\subsection{Formalism}
Although a similar treatment of time-dependent Bogoliubov-de Gennes (BdG) equations has already been used in the context of Majorana fermions,\cite{Cheng} we present our own formulation best suited for the analysis in this paper.
Let us begin with the general time-dependent BCS mean-field Hamiltonian
\begin{equation}
 \hat{\mathcal{H}}(t) = \frac{1}{2}\int\diff\vec{r}\int\diff\vec{r}\,'\, \hat{\Psi}^\dag(\vec{r})h(t) \hat{\Psi}(\vec{r}\,')
 \label{General_expressions:GeneralHamiltonian}
\end{equation}
written in the standard quadratic BdG form. For simplicity, spinor indices have been skipped. The field-operators $\hat{\Psi}(\vec{r})$ are Nambu spinors satisfying the Majorana condition,
\begin{equation}
C_{j,k} \hat{\Psi}^\dag_k(\vec{r}) = \hat{\Psi}_j(\vec{r}),  
\label{General_expressions:MajCond}
\end{equation}
where $C$ is the (unitary) spinor part of the (antiunitary) charge conjugation operator $\Xi = C\mathcal{K}$ with $\mathcal{K}$ denoting complex conjugation. Throughout this paper, we assume that $\Xi^2=+\mathbbm{1}$ restricting the analysis to superconductors of class $D$. Due to its internal redundancy $\hat{\Psi}(\vec{r})$ satisfies the Majorana anticommutation relations,
\begin{subequations}
\begin{align}
    \left\{\hat{\Psi}_j(\vec{r}),\hat{\Psi}_k(\vec{r}\,')\right\} &= C_{j,k}\delta(\vec{r}-\vec{r}\,'), \\ 
    \left\{\hat{\Psi}_j(\vec{r}),\hat{\Psi}^\dag_k(\vec{r}\,')\right\} &= \delta_{j,k}\delta(\vec{r}-\vec{r}\,'),
\end{align}
\end{subequations}
as opposed to those of ordinary fermions.
Additionally, charge conjugation symmetry imposes the constraint $\acom{h(t)}{\Xi} = 0$ making the instantaneous spectrum symmetric about zero energy.

Note that the ansatz (\ref{General_expressions:GeneralHamiltonian}) only allows for the coupling of the system to a classical field. Apart from that, the time-dependence of $\hat{\mathcal{H}}(t)$ is not further specified throughout this section. To be concrete, one may imagine that vortices of a two-dimensional $p_x+ip_y$ superconductor\cite{Ivanov} or domain walls in a quantum wire are manipulated, \eg by the local tuning of external gates\cite{vonOppen} as illustrated in \figref{GeneralSystem}(a). 

Let us introduce the instantaneous eigenstates $\ket{\phi_n(t)}$ of the BdG Hamiltonian satisfying 
\begin{equation}
 h(t)\ket{\phi_n(t)} = E_n(t)\ket{\phi_n(t)},
\label{General_expressions:InstantEigenStates}
\end{equation} 
which are chosen to be continuous as a function of $t$. To fix the relative phase of the instantaneous eigenstates at different times we impose the parallel transport condition
\begin{equation}
 \braket{\phi_n(t)|\partial_t \phi_n(t)} = 0.
\label{General_expressions:ParallelTransport}
\end{equation}
This allows us to define the corresponding instantaneous BdG operators
\begin{equation}
\hat{d}_n(t):=\int\mathrm{d}\vec{r}\,\phi_n^\dag(\vec{r},t)\hat{\Psi}(\vec{r}).
\label{General_expressions:instBdGOp}
\end{equation}
Both the field operators $\hat{\Psi}(\vec{r})$ and the eigenfunctions $\phi_n(\vec{r},t)$ have spinor structure and hence summation over spinor components is implied. Physically, these operators correspond to the annihilation of a particle in one of the instantaneous eigenstates at a given time $t$. They constitute the central objects of our analysis, since all physical quantities to be calculated in the following can be written in terms of $\hat{d}_n(t)$ and $\hat{d}_n^\dag(t)$.
It is straightforward to show that $\hat{d}^\dag_n(t) = \hat{d}_{\bar{n}}(t)$ and
\begin{equation}
 \acom{\hat{d}_n(t)}{\hat{d}_m(t)} = \delta_{n,\overline{m}}, \quad \acom{\hat{d}_n(t)}{\hat{d}^\dag_m(t)} = \delta_{n,m},
\label{General_expressions:instBdGcom}
\end{equation}
where $\bar{n}$ is a short hand notation for the charge conjugate of state $n$ ($E_{\bar{n}}=-E_n$).

Suppose that diagonalizing $h(t)$ yields $M$ pairs of quasi-zero energy subgap states $\{\ket{\phi_{0j}(t)},\ket{\phi_{\cbar{0j}}(t)}\}$ with $j=1,2,\dots,M$. Denoting the corresponding BdG operators by $\hat{d}_{0j}(t)$, we can define $2M$ time-dependent MBS operators 
\begin{subequations}
\begin{align}
  \hat{\gamma}_{2j-1}(t)&:=\frac{1}{\sqrt{2}}\left(\hat{d}_{0j}(t) + \hat{d}_{0j}^\dag(t)\right), \\ 
  \hat{\gamma}_{2j}(t)&:=\frac{1}{\sqrt{2}i}\left(\hat{d}_{0j}(t) - \hat{d}_{0j}^\dag(t)\right),
\end{align}
\label{General_expressions:MajoranaFermion}
\end{subequations}
which satisfy by construction 
\begin{equation}
\hat{\gamma}^\dag_j(t) = \hat{\gamma}_j(t) \quad\text{and}\quad \acom{\hat{\gamma}_{i}(t)}{\hat{\gamma}_{j}(t)} = \delta_{i,j}.
\end{equation}
The energies $E_{0j}$ typically scale exponentially with the distance between different MBS (see \eg \refcite{ChengMeng}). \equref{General_expressions:MajoranaFermion} makes clear that a Majorana mode is, in contrast to the ordinary fermions $\hat{d}_n$, generally\footnote{$E_{0j}$ can have isolated zeros as a function of the spatial separation of the MBS.} not an exact eigenstate of the Hamiltonian of a finite system.
We emphasize that for $M>1$ one is left with the additional task of finding the correct superpositions of the $\hat{\gamma}_i$ such that the associated Majorana wavefunctions are spatially localized. However, for $M=1$, \equref{General_expressions:MajoranaFermion} is already sufficient. Note that the parallel transport condition (\ref{General_expressions:ParallelTransport}) is automatically satisfied by the wavefunctions $\ket{\phi_{\gamma_j}}$ of the MBS due to the self-conjugate property $\Xi\ket{\phi_{\gamma_j}}=\ket{\phi_{\gamma_j}}$.

Using the anticommutation relations (\ref{General_expressions:instBdGcom}) it is easy to show that the Heisenberg equations of motion for the instantaneous BdG operators read
\begin{equation}
 i\der{t} \hat{d}^H_n(t) = \sum_m h_{n,m}'(t) \hat{d}^H_m(t), \label{General_expressions:Heisenberg3}
\end{equation} 
where 
\begin{equation}
 \hat{d}^H(t)=\hat{\mathcal{U}}^\dag(t)\hat{d}_n(t)\hat{\mathcal{U}}(t), \quad \hat{\mathcal{U}}(t) =\mathcal{T}e^{-i\int_{t_0}^t \diff t \, \hat{\mathcal{H}}(t')},
\end{equation} 
are the instantaneous BdG operators in the Heisenberg picture and
\begin{equation}
 h'_{n,m}(t) = E_n(t)\delta_{n,m} - \mathcal{M}_{n,m}(t) \label{General_expressions:MovingFrameHamiltonian}
\end{equation} 
with $\mathcal{M}_{n,m}(t)=i\braket{\phi_n(t)|\partial_t\phi_m(t)}$ has been introduced. The summation in \equref{General_expressions:Heisenberg3} includes all instantaneous eigenstates with both positive and negative energy. 
Note that $h'$ is the ``moving frame'' Hamiltonian of $h$, which is well-known\cite{Berry2} from the study of non-adiabatic quantum mechanics. Non-vanishing values of $\mathcal{M}_{n,m}$ are due to the time-dependence of the basis states in \equref{General_expressions:instBdGOp}. For the particularly important case of $n=0j$ and $m$ referring to a state above the gap (as well as $n \leftrightarrow m$), these matrix elements give rise to transitions from the ground state manifold, \ie the topological qubit(s), to excited states. This kind of non-adiabatic processes represents the major focus of our analysis.

The formal solution of the Heisenberg equation (\ref{General_expressions:Heisenberg3}) is given by
\begin{equation}
 \hat{d}^H_n(t) = \sum_m u_{n,m}(t) \hat{d}_m(t_0), \label{General_expressions:HeisenSol1}
\end{equation} 
where the time-ordered matrix exponential
\begin{equation}
 u_{n,m}(t) := \left[\mathcal{T} \exp\left(-i\int_{t_0}^t \diff t' \, h'(t')\right) \right]_{n,m} \label{General_expressions:TimeEvolutionAdiabaticBasis}
\end{equation}
has been defined. In the following we will use \equref{General_expressions:HeisenSol1} to express the non-adiabatic time-evolution of a topological qubit in terms of the matrix elements $u_{n,m}(t)$.

\subsection{Qubit quantities}
\label{Qubitquantities}
For the remainder of this paper, the analysis is restricted to a single topological qubit, \ie four Majorana modes, $\gamma_1,\dots,\gamma_4$, paired into ordinary fermions according to
\begin{subequations}
\begin{align}
 \hat{d}_{01}(t) &= \frac{1}{\sqrt{2}}\left(\hat{\gamma}_1(t)+i\hat{\gamma}_2(t)\right), \\
 \hat{d}_{02}(t) &= \frac{1}{\sqrt{2}}\left(\hat{\gamma}_3(t)+i\hat{\gamma}_4(t)\right).
\end{align} 
\end{subequations}
Further MBS may be present in the sample, but are assumed to be inert and sufficiently far away to be safely neglected. The many-body ground state wavefunctions are defined as usual,
\begin{align}
 \ket{n_1\,n_2\,(t)} := \left(\hat{d}_{01}^\dag(t)\right)^{n_1}\left(\hat{d}_{02}^\dag(t)\right)^{n_2}\ket{00(t)},
\end{align}
where $\ket{00(t)}$ denotes the vacuum of the fermions $\hat{d}_{0j}(t)$, $j=1,2$. Note that, in the present case, the operators and hence the states are time-dependent. Without loss of generality, we take the even fermion parity sector $\{\ket{00(t)},\ket{11(t)}\}$ to form the logical basis of the qubit and assume that it was at the initial time $t_0$ prepared in a pure state within this subspace. In addition, the initial density matrix $\hat{\rho}(t_0)$ is taken to be diagonal in the occupation number basis with respect to the fermions $\hat{d}_n$ of the continuum.

\begin{figure}[b]
\begin{center}
\includegraphics[width=\linewidth]{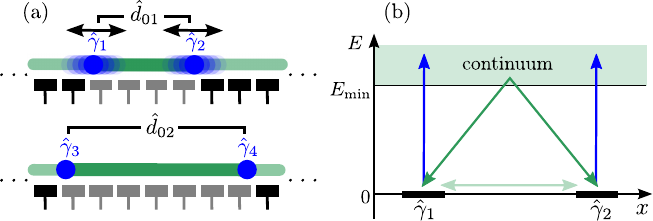}
\caption{(Color online) (a) One example for the origin of the time-dependence of the BdG Hamiltonian in \equref{General_expressions:GeneralHamiltonian}. The position of the Majorana modes is modified via a ``keyboard'' of gates as suggested in \refcite{vonOppen}. Here the situation is shown where only the MBS belonging to the fermion $\hat{d}_{01}$ are in motion. This is the setup we will focus on in \secsref{Limitations}{UTQC}. (b) Illustration of adiabatic (light green arrow) and non-adiabatic (dark green) MBS-MBS as well as MBS-continuum (blue) processes.}
\label{GeneralSystem}
\end{center}
\end{figure}

For simplicity, let us assume that the MBS $\gamma_3$, $\gamma_4$ are both spatially fixed, 
decoupled from $\gamma_1$, $\gamma_2$ and from each other. Thus, their role is reduced to 
providing the proper Hilbert space for the qubit. This means
\begin{equation}
 u_{\gamma_k,n}(t) = \delta_{\gamma_k,n}
 \label{General_expressions:TimeEvAssum}
\end{equation}
for $k=3,4$ (see \figref{GeneralSystem}(a) for an illustration in the system of locally gated nanowires). Since only the Majorana modes $\gamma_1$ and $\gamma_2$ belonging to one and the same fermion contribute to transitions, this situation will be referred to as ``intrafermionic motion'' in the following.
The analysis is readily generalized to the situation, where all four MBS are moving simultaneously, however, the results do not convey additional physical insights. 

\subsubsection{Single fermion parity}

The first quantity we use to describe the dynamics of the topological qubit is the parity of the Dirac fermion $d_{01}$ formed by $\gamma_1$ and $\gamma_2$, 
which is defined by
\begin{equation}
 \hat{\mathcal{P}}_{01}(t):=\unitmatrix-2\hat{d}_{01}^\dag(t)\hat{d}^{\phantom{\dag}}_{01}(t)=2i\hat{\gamma}_{2}(t)\hat{\gamma}_{1}(t).\label{General_expressions:SingleParityDef}
\end{equation} 
 The fermion parity constitutes a frequently used\cite{Cheng,Goldstein} observable to describe the fidelity of the topological memory. We emphasize that, regarding the qubit, a change in the expectation value of $\hat{\mathcal{P}}_{01}(t)$ can in general be due to two distinct processes, namely $\ket{00} \leftrightarrow \ket{10}$ (leaving the subspace of the logical two level system and exciting a quasiparticle in the continuum) and $\ket{00} \leftrightarrow \ket{11}$ (bit-flip error within the logical subspace). In the present case of intrafermionic motion (uncoupled $\gamma_3$ and $\gamma_4$), only errors of the former type can occur.

Using \equref{General_expressions:HeisenSol1}, $\hat{d}^\dag_n = \hat{d}_{\bar{n}}$ and $u_{\overline{n},\overline{m}} = u_{n,m}^*$ due to charge conjugation symmetry as well as the unitarity of $u$, one can write 
\begin{align}
\begin{split}
&\braket{\hat{\mathcal{P}}_{01}(t)}_{t} = \braket{\hat{\mathcal{P}}_{01}(t_0)}_{t_0} \\
&\qquad\times \left(u_{\gamma_1,\gamma_1}(t)u_{\gamma_2,\gamma_2}(t)-u_{\gamma_1,\gamma_2}(t)u_{\gamma_2,\gamma_1}(t) \right)\\
&\qquad+ 2\sum_{n>0} \braket{\hat{\mathcal{P}}_n(t_0)}_{t_0}\Im{u^*_{\gamma_2,n}(t)\,u_{\gamma_1,n}(t)}.
\label{General_expressions:ParityMBSIntra}
\end{split}
\end{align} 
To restate this result in a more explicit form, we now apply time-dependent perturbation theory treating the coupling matrix $\mathcal{M}$ in \equref{General_expressions:MovingFrameHamiltonian} as the perturbation. For the moment let us additionally assume that $\gamma_1$ and $\gamma_2$ are sufficiently separated such that the energy splitting $E_{01}$ and the ground state Berry phases can be neglected. Then second order perturbation theory yields 
\begin{widetext}
\begin{align}
\begin{split}
 \braket{\hat{\mathcal{P}}_{01}(t)}_t &= \braket{\hat{\mathcal{P}}_{01}(t_0)}_{t_0} \left(1-  \sum_{j=1,2} \sum_{n>0} \left| \int_{t_0}^t\mathrm{d}t' \mathcal{M}_{\gamma_j,n}(t') e^{-i \int_{t_0}^{t'}\hspace{-0.2em}\mathrm{d}t_1 E_n(t_1)}  \right|^2 \right) \\
&+ 2\sum_{n>0} \braket{\hat{\mathcal{P}}_n(t_0)}_{t_0}\Im{\left(\int_{t_0}^t\mathrm{d}t' \mathcal{M}_{\gamma_1,n}(t') e^{-i \int_{t_0}^{t'}\hspace{-0.2em}\mathrm{d}t_1 E_n(t_1)}\right) \cdot\left(\gamma_1 \rightarrow \gamma_2\right)^*} + \order{\mathcal{M}^3}.
\label{General_expressions:PertFermPar1}
\end{split}
\end{align}
\end{widetext}
As schematically shown in \figref{GeneralSystem}(b), one can in principle distinguish between two types of processes contributing to the time-evolution of the topological qubit. The MBS can either couple locally to the continuum states (blue arrows) or communicate with each other. The latter type of processes contains both direct MBS-MBS tunneling (light green arrows) and non-adiabatic corrections involving virtual states in the continuum as indicated by the dark green arrows. 

In the results presented above for the single fermion parity, we can identify both contributions. The first term in the second line of \equref{General_expressions:ParityMBSIntra} and the first line of \equref{General_expressions:PertFermPar1} are (to leading order) local, whereas all remaining contributions are purely non-local and describe correlation effects between different MBS. Note that, when neglecting the non-local terms, the result (\ref{General_expressions:PertFermPar1}) for the fermion parity reduces to the expression obtained in \refcite{Goldstein}. 
   
\subsubsection{Off-diagonal component of the density matrix}
For a more refined picture of the time-evolution of the qubit let us investigate the off-diagonal matrix element of its reduced density matrix, \ie
\begin{equation}
\rho^Q_{01}(t) := \braket{00(t)|\hat{\rho}^Q(t)|11(t)} = \braket{\hat{d}^\dag_{01}(t) \hat{d}^\dag_{02}(t)}_t
\label{General_expressions:DensityExpec}
\end{equation}
with $\hat{\rho}^Q(t) = \ptrace{C}{\hat{\rho}(t)}$, where $\ptrace{C}{\cdot}$ stands for the partial trace taken over the continuum states above the gap. 
We emphasize that $\rho^Q_{01}(t)$ is relevant for two reasons: Firstly, the reduction of its magnitude describes decoherence. Note that, in the case of intrafermionic motion, the main decoherence mechanism is leaving the logical Hilbert space of the qubit. Secondly, the change of the phase of $\rho^Q_{01}(t)$ means that a phase gate can be performed by mutual motion of the two MBS. Generating a time-dependent phase of $\rho^Q_{01}(t)$ by non-adiabatic effects may provide alternative routes for implementing phase gates, which are crucial\cite{Nayak} for realizing universal quantum computation (see \secref{UTQC}).

Due to the assumption of intrafermionic motion, we have $u_{\overline{02},n}=\delta_{\overline{02},n}$ and the calculation becomes particularly straightforward:
\begin{align}
  \rho^Q_{01}(t) &= \sum_{n} u_{\overline{01},n}(t) \braket{\hat{d}_{n}(t_0) \hat{d}^\dag_{02}(t_0)}_{t_0}\\
		 &= \rho^Q_{01}(t_0) \cdot u_{\overline{01},\overline{01}}(t,t_0). \label{General_expressions:DecIntra}
\end{align}
Despite its simplicity, it is instructive to restate \equref{General_expressions:DecIntra} in the Majorana basis in the form 
\begin{equation}
 \rho^Q_{01}(t) = \rho^Q_{01}(t_0) \times (c(t) + i\, s(t)) 
\label{ch:General_expressions:Dec2MFMoving}
\end{equation}
with the real-valued\footnote{Particle hole symmetry implies $u_{\overline{n},\overline{m}} = u_{n,m}^*$ and hence $u_{\gamma_j,\gamma_k} \hspace{-0.1em}\in \mathbbm{R}$.} functions
\begin{subequations}
\begin{align}
 c(t) &= \frac{1}{2} (u_{\gamma_1,\gamma_1}(t) + u_{\gamma_2,\gamma_2}(t)), \label{General_expressions:SCfunctionsC}\\
 s(t) &= \frac{1}{2} (u_{\gamma_1,\gamma_2}(t) - u_{\gamma_2,\gamma_1}(t)). \label{General_expressions:SCfunctionsS}
\end{align}\label{General_expressions:SCfunctions}\end{subequations}
This result reveals that non-vanishing matrix elements $u_{\gamma_j,\gamma_k}$ with $j\neq k$ in \equref{General_expressions:SCfunctionsS}, \ie non-local processes, are required to have $s(t)\neq 0$ and are thus essential for generating a time-dependent phase of $\rho^Q_{01}(t)$. Without these processes, the system can only experience decoherence due to the local terms in \equref{General_expressions:SCfunctionsC}.     

Again assuming well-separated MBS, one finds within second order perturbation theory in $\mathcal{M}$
\begin{equation}
\rho^Q_{01}(t) = \rho^Q_{01}(t_0) \cdot \exp(\Gamma(t) + i\varphi(t)) + \order{\mathcal{M}^3},
 \label{General_expressions:DensityExponentNot}
\end{equation} 
where the decoherence function and the accumulated phase are given by
\begin{widetext} 
\begin{subequations}
\begin{align}
  \Gamma(t) = - \frac{1}{2}\sum_{j=1,2}\sum_{n>0} \left| \int_{t_0}^t\mathrm{d}t' \mathcal{M}_{\gamma_j,n}(t') e^{-i \int_{t_0}^{t'}\hspace{-0.2em}\mathrm{d}t_1 E_n(t_1)} \right|^2
 \label{General_expressions:DecoherenceFunction}
\end{align}
and 
\begin{equation}
   \varphi(t) = \sum_{n>0}\left(\int_{t_0}^t\mathrm{d} t'\int_{t_0}^{t'}\mathrm{d} t''\,\Re{\mathcal{M}_{\gamma_2,n}(t')\mathcal{M}_{\gamma_1,n}^*(t'')e^{-i\int_{t''}^{t'}\mathrm{d} t_1E_n(t_1)}} - \left(\gamma_1\leftrightarrow \gamma_2\right)\right),
 \label{General_expressions:PhaseGenerated}
\end{equation} 
\label{General_expressions:PhaseDec}
\end{subequations}
\end{widetext}
respectively. Here the notation $n>0$ indicates that the summation is restricted to positive energy eigenstates of the continuum. The negative energy eigenstates have been replaced by means of the relation $\mathcal{M}_{\overline{n},\overline{m}}=-\mathcal{M}^*_{n,m}$.

As expected from the exact expression (\ref{General_expressions:SCfunctions}), the leading contribution to the decoherence is only due to the local coupling of each of the MBS to the continuum and thus solely depends on the motions of the Majorana modes separately. 
On the contrary, the phase $\varphi(t)$ generated by the process crucially depends on the correlation of the motions of the spatially separated $\gamma_1$ and $\gamma_2$ and can only be present if the coupling of both MBS to the continuum is non-zero.   
As expected, the change of the fermion parity (leaving the logical Hilbert space of the qubit) in \equref{General_expressions:PertFermPar1} is directly related to the decoherence function (\ref{General_expressions:DecoherenceFunction}), if only one of the Majorana bound states is in motion.

\subsection{Effective theory for nearly adiabatic manipulation}
\label{EffectiveTheory}
In this subsection an effective Hamiltonian governing the dynamics within the ground state manifold is derived. Since non-adiabatic effects are treated in a perturbative manner, its validity is limited to the regime of nearly adiabatic processes.
Here we consider the general case of $2M$ Majorana modes and account for finite overlaps between different localized states. 

\begin{figure}[b]
\begin{center}
\includegraphics[width=\linewidth]{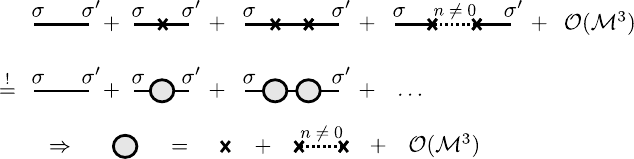}
\caption{Schematic illustration of the procedure for determining the effective ground state Hamiltonian according to \equref{General_expressions:DefiningPropOfEffHam}. Excited states (dashed lines) are absorbed into an effective ground state vertex (circle). For slowly varying energies and coupling matrix elements (crosses) the result in the last line is approximately local in time (see main text).}
\label{EffectiveHamitlonianGraphicalRep}
\end{center}
\end{figure}

To integrate out the continuum states a method taken from \refcite{Hutter,Jose} is applied which we generalize to the case of time-dependent energies $E_n(t)$. 
The basic idea is to find an effective Hamiltonian $h_{\text{eff}}(t)$ that reproduces the exact time-evolution operator of $h'(t)$ in \equref{General_expressions:MovingFrameHamiltonian} within the ground state subspace. In the interaction picture (again taking $h'_1=-\mathcal{M}$ as perturbation), we demand that
\begin{equation}
 P_0\,\mathcal{T}e^{-i\int_{t_0}^t \diff t'\,\tilde{h}'_{1}(t')}P_0 \stackrel{!}{=} \mathcal{T}e^{-i\int_{t_0}^t \diff t'\,\tilde{h}_{\text{eff},1}(t')},
 \label{General_expressions:DefiningPropOfEffHam}
\end{equation} 
where $P_0:=\sum_\sigma \ketbra{\phi_\sigma}$ is the projection operator onto the ground state manifold. In \figref{EffectiveHamitlonianGraphicalRep}, a graphical representation of the expansion of both sides of \equref{General_expressions:DefiningPropOfEffHam} and of the resulting effective interaction Hamiltonian $\tilde{h}_{\text{eff},1}$ within second order in $\mathcal{M}$ is shown. 
Since the second order contribution in the last line of \figref{EffectiveHamitlonianGraphicalRep} has two time-arguments, further approximations are required to obtain an effective Hamiltonian which is local in time. For this purpose, let us assume that both all coupling matrix elements $\mathcal{M}_{\sigma,n}$ and the instantaneous energies $E_n$ vary slowly on the scale set by the gap $E_{\text{min}}$ of the system, or more formally 
\begin{equation}
 \frac{\partial_t E_n(t)}{E^2_\text{min}}\ll 1, \qquad \frac{1}{E_\text{min}}\frac{\partial_t B_{\sigma,n}(t)}{B_{\sigma,n}(t)} \ll 1. \label{General_expressions:NearlyAdiabatic}
\end{equation} 
An expansion up to first order in these small quantities finally yields
\newcommand{\negspace}{\hspace{-0.1em}}
\begin{align}
\begin{split}
&\left(h_{\text{eff}}(t)\right)_{\sigma,\sigma'} \approx E_\sigma\delta_{\sigma,\sigma'} -\mathcal{M}_{\sigma,\sigma'} \\
&\qquad-\sum_{n\neq 0}\Biggl[ \frac{\mathcal{M}_{\sigma,n} \mathcal{M}_{\sigma'\negspace,n}^*}{E_n-\Ebar_{\sigma,\sigma'}} \left(1 - \frac{i\partial_t\left(E_{\sigma}-E_{\sigma'}\right)}{4\left(E_n-\Ebar_{\sigma,\sigma'}\right)^2}\right) \\ 
&\qquad-i\frac{\dot{\mathcal{M}}_{\sigma,n}\mathcal{M}^*_{\sigma'\negspace,n}-\mathcal{M}_{\sigma,n}\dot{\mathcal{M}}^*_{\sigma'\negspace,n}}{2(E_n-\Ebar_{\sigma,\sigma'})^2} \Biggr], \label{General_expressions:EffectiveHamiltonianGeneralResult}
\end{split}
\end{align}
upon introducing $\,\Ebar_{\sigma,\sigma'}:=(E_\sigma+E_{\sigma'})/2$. Here the explicit time-dependence of the matrix elements and of the energies has been omitted for notational convenience.
The first order contributions (first line in \equref{General_expressions:EffectiveHamiltonianGeneralResult}) are simply given by the projection of $h'(t)$ onto the ground state manifold and thus describe adiabatic processes only, whereas the second order terms (second and third line) constitute non-adiabatic corrections incorporating high energy degrees of freedom above the gap in the form of single virtual states. 
Note that the effective Hamiltonian is Hermitian within the present approximations and hence the total parity of the ground state subspace is conserved. 
Interestingly, the last line in \equref{General_expressions:EffectiveHamiltonianGeneralResult} is a sum of terms of the form $i(x\dot{y}-\dot{x}y)/2$, upon choosing $x=\mathcal{M}^*_{\sigma'\negspace,n}/(E_n-\Ebar_{\sigma,\sigma'})$ and $y=\mathcal{M}_{\sigma,n}/(E_n-\Ebar_{\sigma,\sigma'})$. Since $x,y\rightarrow 0$ for $|t|\rightarrow\infty$, the time integral $\int_{-\infty}^\infty\diff{t}\,(x\dot{y}-\dot{x}y)/2$ is given by the area enclosed by the trajectory $(x(t),y(t))$. Consequently, the leading contribution of this term to the time-evolution has a purely geometric interpretation.

For future reference, let us investigate the simplest case with only two MBS ($M=1$). Taking the particle hole symmetry of $h_{\text{eff}}$ into account, we know that $\left(h_{\text{eff}}\right)_{\bar{\sigma},\sigma} = - \left(h_{\text{eff}}\right)_{\sigma,\bar{\sigma}}^*=-\left(h_{\text{eff}}\right)_{\bar{\sigma},\sigma}=0$.
Therefore, the presence of the excited states only leads to a renormalization of the energy splitting of the MBS, \ie the effective Hamiltonian written in the basis $\{\ket{\phi_{\sigma}},\ket{\phi_{\bar{\sigma}}}\}$ assumes the simple form
\begin{equation}
 h_{\text{eff}}(t) \approx (E_{01}(t) + \delta E(t))\tau_z, \label{General_expressions:EffHamGen2MBS}
\end{equation} 
where the energy correction is given by 
\begin{align}
 \begin{split} \delta E(t) &= -\sum_{n\neq 0}\Biggl[\frac{\left|\mathcal{M}_{01,n}\right|^2}{E_n-E_{01}} \\
 & \qquad -i\,\frac{\dot{\mathcal{M}}_{01,n}\mathcal{M}^*_{01,n}-\mathcal{M}_{01,n}\dot{\mathcal{M}}^*_{01,n}}{2\left(E_n-E_{01}\right)^2}\Biggr] \label{General_expressions:EnergyCorr1}                                                                                                                                                                     \end{split} \\
\begin{split} &\approx \sum_{n>0}\Biggl[\frac{2\,\Im{\mathcal{M}_{\gamma_1,n}^* \mathcal{M}_{\gamma_2,n}}}{E_n} \\ 
& \qquad + \frac{\Re{\dot{\mathcal{M}}^*_{\gamma_1,n}\mathcal{M}_{\gamma_2,n}-\mathcal{M}^*_{\gamma_1,n}\dot{\mathcal{M}}_{\gamma_2,n}}}{E_n^2}\Biggr] \label{General_expressions:EnergyCorr2}.\end{split}
\end{align} 
To obtain \equref{General_expressions:EnergyCorr2}, we have neglected all terms $\order{E_{01}/E_n}$, restated the matrix elements in the basis of the localized Majorana wavefunctions and exploited the charge conjugation symmetry one more time to reduce the summation to continuum states with positive energy ($n>0$).  

\section{Non-adiabatic limitations for a quantum wire}
\label{Limitations}
\begin{figure}[b]
\begin{center}
\includegraphics[width=\linewidth]{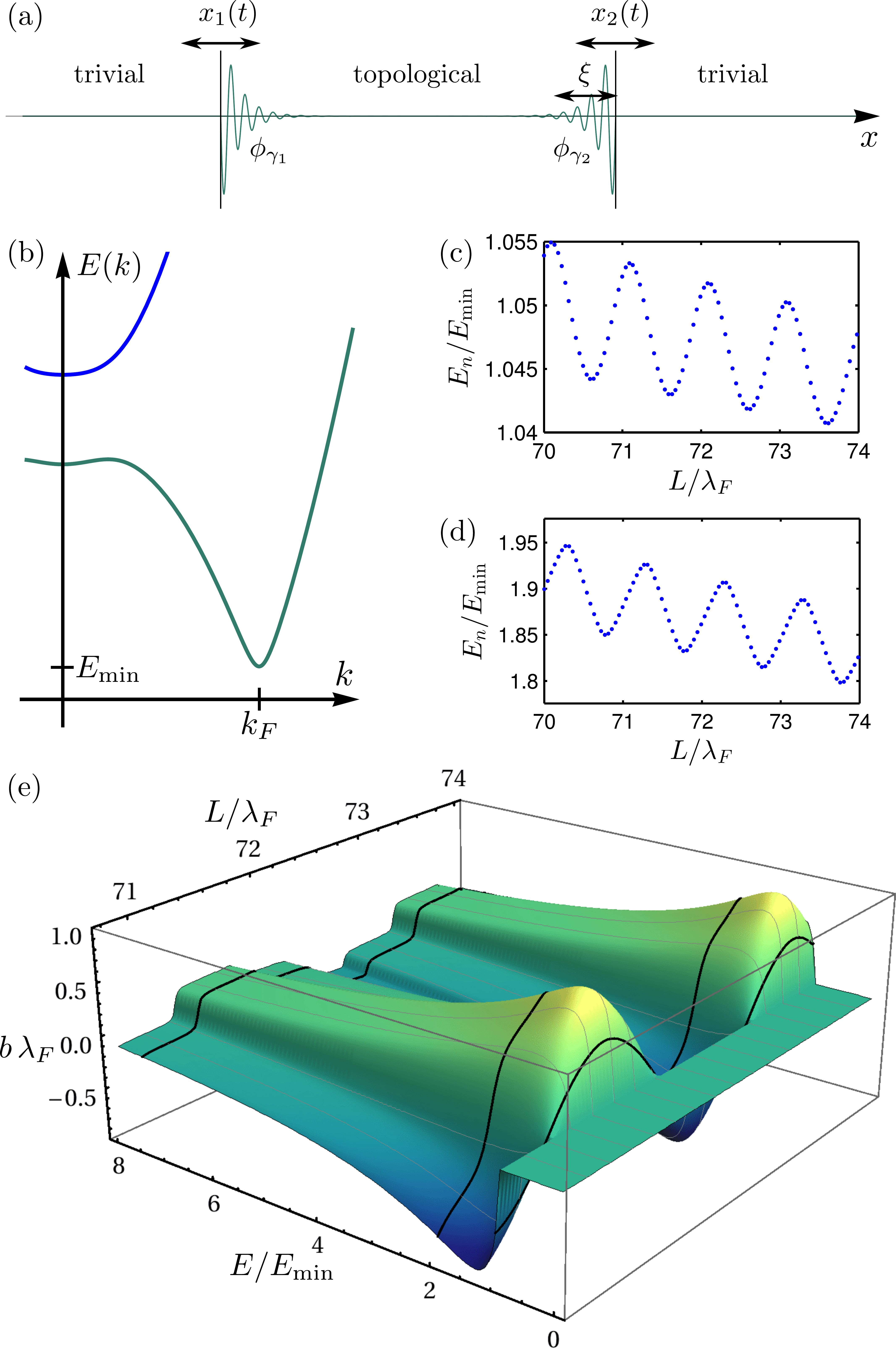}
\caption{(Color online) Due to the assumption of intrafermionic manipulation, it is sufficient to analyze an isolated topological segment (a) of the wire surrounded by a trivial phase. In (b) the excitation spectrum of the BdG Hamiltonian (\ref{Majorana_Fermions:WireBdG}) for large magnetic fields $(B\gg \epsilon_{\text{so}},\Delta)$ is shown. The lower band (green) is effectively described by Kitaev's spinless model (\ref{Quantum_wire:ContKitModel2}). In (c) and (d) we show the $L$-dependence of the instantaneous energies of two different states determined numerically from $h_{\text{NW}}$ for $B/\Delta=5$, $B/\epsilon_{\text{so}}=20$ and taking $\mu=0$ in the topological segment. In (e) the analytical expression (\ref{Quantum_wire:MatrixElementsResults}) for the geometric matrix elements is plotted as a function of the length $L$ of the topological domain and the energy $E$ of the excited state using the same parameters as in (c) and (d).}
\label{Dispersion3DPlot}
\end{center}
\end{figure}
To study the limitations on topological quantum computing caused by non-adiabatic effects, we now apply the general results (\ref{General_expressions:PhaseDec}) and (\ref{General_expressions:PertFermPar1}) for intrafermionic processes to the quantum wire proposal. Since $\gamma_3$ and $\gamma_4$ are assumed to be irrelevant for the qubit dynamics, we can focus on the single topological domain depicted in \figref{Dispersion3DPlot}(a). Restricting the analysis to the case where only the positions $x_1$ and $x_2$ of the domain walls are varied and taking the localization length $\xi$ of the MBS wavefunctions to be much larger than their separation $L=x_2-x_1$, one can write 
\begin{equation}
 \mathcal{M}_{\gamma_j,n}(t) \approx i\,\dot{x}_j(t)\,b_{\gamma_j,n}(L(t)),
\end{equation}   
where the geometric matrix elements are defined by $b_{\gamma_j,n}:=\braket{\phi_{\gamma_j}|\partial_{x_j} \phi_n}$. The time-dependence of $x_1$ and $x_2$ may generally be due to random gate fluctuations or the consequence of intentional successive tuning of local gate voltages for the purpose of information processing. 
  
Focusing first on the simplest case $\dot{x}_1=0$, \equsref{General_expressions:PhaseDec}{General_expressions:PertFermPar1} are solely determined by $I_2$, where
\begin{equation}
 I_j:=\sum_{n>0} \left| \int_{t_0}^t\diff t'\dot{x}_j(t')\,b_{\gamma_j,n}(L(t')) e^{-i\int_{t_0}^{t'}\hspace{-0.2em}\diff t_1 E_n(t_1)} \right|^2  \ .\label{Quantum_wire:CentralIntegral}
\end{equation}
The decoherence function reads $\Gamma=-I_2/2$ and $I_2$ represents, at the same time, the probability of changing the single fermion parity. 
To evaluate $I_2$, we first need to calculate the geometric matrix elements for the specific setup under discussion. The BdG Hamiltonian of a nanowire proximity coupled to an $s$-wave superconductor is given by\cite{Oreg, Lutchyn}
\begin{equation}
 h_{\text{NW}} = \left(\frac{p^2}{2m} - \mu(x) + u p \,\sigma_y\right)\tau_z + B \sigma_z - \Delta \tau_x, \label{Majorana_Fermions:WireBdG}
\end{equation} 
where the Pauli matrices $\sigma_j$ and $\tau_j$ act on spin and particle hole space, respectively. Here $m$ denotes the effective mass, $u$ the spin-orbit coupling strength, $\Delta$ the induced pairing potential and $B$ the Zeeman energy. 

\subsection{Large magnetic fields and infinite potential well}
To obtain analytical results, we focus on the limit where the magnetic field is the largest energy scale of the topological segment, \ie $B\gg \epsilon_{\text{so}},\Delta$ with $\epsilon_{\text{so}}=mu^2/2$ representing the spin-orbit energy. Under these assumptions one can project \equref{Majorana_Fermions:WireBdG} onto its lower band (green line in \figref{Dispersion3DPlot}(b)) yielding\cite{vonOppen} the (continuum limit of the) Kitaev model,\cite{Kitaev}
\begin{equation}
 h_{\text{Kit}} = \left(\frac{p^2}{2m} - \mu_e(x) \right)\tau_z - v_e p \,\tau_y, \label{Quantum_wire:ContKitModel2}
\end{equation} 
as an effective low energy theory, where $\mu_e = \mu + B$ and $v_e = u\Delta/B$. 

Let us first consider an infinite potential well, \ie we need to find the eigenstates $\phi$ of \equref{Quantum_wire:ContKitModel2} with $\mu_e(x)=\mu_1>0$ subject to the constraint $\phi(x_1)=\phi(x_2)=0$.
One can derive an analytical expression (see \appref{Ap:MatrixElements}) for the geometric matrix elements taking the energy $E$ of the continuum state $\phi_n$, $n>0$, as a continuous quantity and considering the limit $L \gg \xi,\lambda_F$, where $\lambda_F=2\pi/k_F$ denotes the Fermi wavelength.
Here we simply show a plot of the result as a function of the system length $L$ and energy $E$ (see \figref{Dispersion3DPlot}(e)) and discuss its relevant properties.
Most importantly, the matrix elements exhibit an oscillatory behavior as a function of $L$ with periodicity of $2\lambda_F$ besides the expected slowly varying envelope function $\propto \sqrt{L}$. The latter can hardly be seen in \figref{Dispersion3DPlot}(e).   
Directly above the gap (rightmost black curve in \figref{Dispersion3DPlot}(e)) the geometric matrix elements are nearly sinusoidal functions of $L$. However, already at $E/E_{\text{min}}=2$ (middle black curve) the sinusoidal shape is significantly deformed. For even higher energies (leftmost black line) smaller than but comparable with $\mu_1$ the matrix elements exhibit a rather step function like $L$-dependence. 
The amplitude of the $L$-oscillations of the matrix elements approaches zero in a non-analytic way ($\propto \sqrt{E-E_{\text{min}}}$) as $E\rightarrow E_\text{min}^+$, reaches its maximum at approximately $E/E_{\text{min}}=1.7$ for the parameters used in \figref{Dispersion3DPlot}(e) and then decays monotonically for larger energies.

In \figref{Dispersion3DPlot}(c) and (d) the numerically determined $L$-dependence of the instantaneous energy of a state directly above the gap and at a higher energy is shown. We again encounter an oscillatory contribution on top of the usual $1/L^2$-decay (this time with periodicity $\lambda_F$). The ratio of its amplitude to the mean energy value increases with energy and decays as $\propto 1/L$.

The oscillation of the energies, which on its own can give rise to transitions,\cite{TienGordon} and the complicated functional form of $b_{\gamma_2}(E,L)$ in \equref{Quantum_wire:MatrixElementsResults} makes a quantitative analytical evaluation of $I_2$ in \equref{Quantum_wire:CentralIntegral} 
very difficult. Nonetheless, we can extract the qualitative behavior of the system from the results presented above. This is achieved as follows.
From \equref{Quantum_wire:CentralIntegral} it is immediately clear that non-adiabatic effects will contribute significantly when the geometric matrix elements have non-vanishing spectral weight for frequencies $\omega \gtrsim E_{\text{min}}$. Due to the sinusoidal behavior of $b_{\gamma_2}(L)$ for low energies, we conclude that  
\begin{equation}
 v_c = \frac{2E_{\text{min}}}{k_F} \approx 2v_e \approx 2u\frac{\Delta}{B}  \label{Quantum_wire:CriticalConcrete1}
\end{equation} 
is the critical velocity scale separating the regimes of nearly adiabatic manipulation ($\dot{L}\ll v_c$) and $\dot{L} \gtrsim v_c$, where non-adiabatic effects render the qubit unstable.

For retaining the topological protection during a braiding process the MBS have to be separated by a distance which is at least a few times larger than their spatial decay length $\xi \approx (m v_e)^{-1}$. Consequently, the Majorana modes are to be transported over a distance $\Delta x\approx \alpha/(mv_e)$ with $\alpha \approx 10-100$ depending on the braiding operation to be realized. According to \equref{Quantum_wire:CriticalConcrete1}, adiabaticity requires the braiding time $T_\text{b}$ to satisfy
\begin{equation}
 T_\text{b} \gg \frac{\Delta x}{v_c} \approx \frac{\alpha}{2mv_e^2}\approx   \frac{\alpha}{4\epsilon_{\text{so}}}\left(\frac{B}{\Delta}\right)^2. \label{Quantum_wire:TimeScale1}
\end{equation}      
We emphasize the difference to the ``standard guess'' of $T_\text{b} \gg 1/E_{\text{min}}$ which is much less restrictive since $\Delta x k_F \gg \Delta x/\xi \gg 1$ in the considered limit of large magnetic fields.

Applying recent experimental data\cite{Mourik} and assuming an order of magnitude difference between the critical braiding time and the lower bound for $T_b$, one finds $T_\text{b} > 10^{-8}\,\mathrm{s}$ for $\alpha\approx 50$. Note that this is already of the same order as the upper bound $T_b<10^{-7}-10^{-8}\,\mathrm{s}$ due to quasiparticle poisoning for the system under discussion.\cite{LossPoisoning} We conclude that non-adiabatic effects due to the presence of degrees of freedom above the gap are not only irrelevant corrections of purely academic interest, but may provide serious challenges for the realizability of topological quantum computing. Note that the authors of \refcite{LossPoisoning} came to a similar conclusion, however, using $T_\text{b} \gg 1/E_{\text{min}}$ and assuming a much smaller gap than that reported in \refcite{Mourik}.

To discuss the typical scaling behavior of the non-adiabatic corrections in the regime $\dot{L}\ll v_c$, let us investigate the prototypal trajectory
\begin{equation}
  L(t) = L(0) + \frac{\Delta x}{\pi}\arctan(t/\tau). \label{Quantum_wire:Trajectory}
\end{equation} 
Assuming that the physics is mainly described by the contribution of the states directly above the gap, we can use the sinusoidal matrix elements $b_{\gamma_2}(L)$. Furthermore, let us neglect the $L$-dependence of the instantaneous energies which can always be justified by choosing $L$ sufficiently large. 
Upon defining $\beta:=\Delta x/(2\lambda_F)$, which measures the number of oscillations of the geometric matrix elements during the trajectory, and $v_{\mu}:=\sqrt{2\mu_1/m}$ one finds 
\begin{align}
\begin{split}
 I_2 \sim \frac{1}{4\sqrt{2\pi}} &\sqrt{\mu_1\tau}\sqrt{\frac{v_e}{v_{\mu}}} \left(\frac{\Delta x/\tau}{v_e}\right)^2  \\ &\times\left(\frac{(2E_{\text{min}}\tau)^\beta}{\Gamma(1+\beta)}\right)^2 e^{-2E_{\text{min}}\tau}, \label{Quantum_wire:ResultProbability}
\end{split}
\end{align}
as $E_{\text{min}}\tau,\,v_\mu/v_e\rightarrow\infty$, \ie for adiabatically slow manipulation and large magnetic fields.
Note that the integral depends on $\Delta x$ and $\tau$ independently, which physically stems from the fact that the system not only has an intrinsic time scale, $E_{\text{min}}$,  but also a length scale $\lambda_F$.

Most importantly, we have found that the non-adiabatic corrections decay exponentially as a function of $E_{\text{min}}\tau$. However, in the adiabatic limit 
($\beta < E_{\text{min}}\tau$) the 
$\beta$-dependent prefactor in \equref{Quantum_wire:ResultProbability} is exponentially large in absolute terms but sub-leading with respect to $e^{-2E_{\text{min}}\tau}$. 
We emphasize that the exponential scaling behavior is directly related to the realistic choice of an analytic protocol $L(t)$.
The non-adiabatic corrections to the fermion parity reported in \refcite{Cheng} vanish only algebraically as the braiding velocity approaches zero, since the authors assumed a discontinuous velocity profile.

These results indicate that it may be favorable to keep the length $L$ of the topological domain constant during a braiding process, \ie $\dot{x}_1=\dot{x}_2$.
In this case both $I_1$ and $I_2$ are required to estimate the decoherence effects in \equsref{General_expressions:DecoherenceFunction}{General_expressions:PertFermPar1}.
Since both the geometric matrix elements and the instantaneous energies are constant, the evaluation of $I_j$ is now readily performed analytically.  
Thus, we obtain for the trajectory (\ref{Quantum_wire:Trajectory}) in the limit $v_\mu/v_e\rightarrow\infty$
\begin{equation}
 I_1=I_2 \sim \frac{1}{2\sqrt{2\pi}} \sqrt{\mu_1\tau}\sqrt{\frac{v_e}{v_{\mu}}} \left(\frac{\Delta x/\tau}{v_e}\right)^2 \,e^{-2E_{\text{min}}\tau} \ .\label{Translation_of:ResultForIj}
\end{equation}  
The second line of \equref{General_expressions:PertFermPar1} is expected to be negligibly small. 
Note that \equref{Translation_of:ResultForIj} is independent of $L$ which is a consequence of the summation over the different states of the continuum. This result implies that parity errors and decoherence effects due to non-adiabatic processes are exponentially suppressed as $e^{-2E_{\text{min}}\tau}$ and consequently lead to the weaker constraint $\tau\gg\tau_c =1/E_{\text{min}}$ for adiabatic quantum computation, as compared to the case of one moving MBS. Naturally, these arguments are only valid as long as the wire is sufficiently clean such that the mean free path is larger than the wire length $L$, which is assumed throughout this paper.

\subsection{Large spin-orbit coupling}
So far we have analyzed the wire Hamiltonian (\ref{Majorana_Fermions:WireBdG}) only in the limit of strong magnetic fields ($B\gg \epsilon_{\text{so}},\Delta$). 
However, also the regime where the spin-orbit coupling at the Fermi level is much larger than the magnetic field and the proximity gap, \ie $\epsilon_{\text{so}}\gg B,\Delta$, is appropriate for engineering MBS. For $\mu=0$ the gap at the Fermi wavevector $k_F \approx 2mu$ is approximately given by $\Delta$. Sufficiently far away from the topological phase transition (explicitly for $B>2\Delta$) the minimal gap in the system occurs at $k_F$. 

We have shown by diagonalizing $h_\text{NW}(L)$ numerically that, also in this limit, the geometric matrix elements $b_{\gamma_j,n}$ are sinusoidal functions of $L$ with periodicity $2\lambda_F$ in the vicinity of the gap and become increasingly deformed for higher energies similarly to \figref{Dispersion3DPlot}(e). The decay length of the Majorana wavefunction is given by $u/\Delta$ in the present regime of the nanowire\cite{Loss} and thus the critical velocity as well as the associated restriction on $T_{\text{b}}$ read
\begin{equation}
 v_c = \frac{2E_{\text{min}}}{k_F} \approx u\frac{\Delta}{2\epsilon_{\text{so}}}, \quad T_\text{b} \gg \frac{\Delta x}{v_c} \approx \frac{2\alpha}{\epsilon_{\text{so}}}\left(\frac{\epsilon_{\text{so}}}{\Delta}\right)^2,
\end{equation} 
respectively. Note that this result has the same structure as \equref{Quantum_wire:TimeScale1} obtained in the limit of large $B$ upon replacing the ratio $B/\Delta$ by $\epsilon_{\text{so}}/\Delta$ and hence the lower boundaries for $T_{\text{b}}$ are expected to be of the same order in both regimes of the nanowire.

\section{Quantum computing using non-adiabatic effects}
\label{UTQC}
In the previous section, non-adiabatic effects have been treated solely as a drawback for topological quantum computing. However, they may also be generated on purpose in order to construct additional gate operations. 
In the following we analyze how non-adiabatic processes can be used to realize a phase gate, $\exp(i\varphi\sigma_z/2)$, and thus, as a special case, a $\pi/8$ gate ($\varphi=\pi/4$) -- the missing single qubit gate for universal quantum computation.\cite{Nayak}

By combining \equsref{General_expressions:DecIntra}{General_expressions:EffHamGen2MBS} we find that the phase $\varphi$ accumulated during a manipulation of the system can be written as 
\begin{equation}
 \varphi = \int_{-\infty}^\infty \diff t'\, (E_{01}(t')+\delta E(t')) \label{Translation_of:PhaseOfRho}
\end{equation} 
in the regime of nearly adiabatic manipulation defined by \equref{General_expressions:NearlyAdiabatic}. This is the limit of interest since a proper gate operation conserves the fermion parity and the coherence of the qubit. Note that, by construction, \equref{Translation_of:PhaseOfRho} can be retrieved directly from the more general expression (\ref{General_expressions:PhaseGenerated}) by a formal expansion in the small parameters in \equref{General_expressions:NearlyAdiabatic}.

The most obvious way for controllably generating a phase $\varphi$ would be to use the tunnel splitting $E_{01}$, \ie to bring the MBS close together for a certain amount of time.
This approach has already been analyzed intensively in past.\cite{PhaseGate,FreedmanNayakPhase} Here we ask whether it is possible to use the non-adiabatic energy splitting $\delta E$ to accumulate a well-defined phase without having direct overlap of the Majorana wavefunctions.

For this purpose let us rewrite the leading term of the energy correction (\ref{General_expressions:EnergyCorr1}) as
\begin{equation}
 \delta E(t) = -\braket{\partial_t\phi_{01}(t)|G|\partial_t\phi_{01}(t)}\left(\negspace 1+\order{\frac{E_{01}}{E_n}} \right), \label{Translation_of:dEThird}
\end{equation} 
where we have introduced the Green's function
\begin{equation}
 G = \sum_n \frac{\ket{\phi_n(t)}\bra{\phi_n(t)}}{E_n} \label{Translation_of:GreensFunctionDefinition}
\end{equation} 
of the BdG Hamiltonian $h(t)$. To obtain the summation over all states in \equref{Translation_of:GreensFunctionDefinition} we have used the parallel transport phase convention (\ref{General_expressions:ParallelTransport}) and $\braket{\phi_{01}|\partial_t \phi_{\cbar{01}}}=0$. 
Since the non-adiabatic accumulation of a phase is a correlation effect between spatially separated MBS (see \secref{Qubitquantities}), the appearance of a Green's function is quite natural. 
By calculating the Green's function $G$ for the case of the infinite potential well described by the Kitaev model (\ref{Quantum_wire:ContKitModel2}) we have shown that the energy correction $\delta E$ in \equref{Translation_of:dEThird} is exponentially small ($e^{-L/\xi}$) in the separation $L$ of the MBS. 
This was to be expected since $G$ describes the (zero frequency) propagation in a gapped system and thus decays exponentially in space. Taking into account the zero energy boundary modes does not change this conclusion. 
Note that $\mathcal{M}_{\gamma_1,n}\mathcal{M}_{\gamma_2,n}^*\in i\,\mathbbm{R}$ for the Kitaev model and hence the second line of \equref{General_expressions:EnergyCorr2} vanishes entirely.
Consequently, also the next order contribution to the non-adiabatic energy splitting (second line of \equref{General_expressions:EnergyCorr1}) is at least of linear order in $E_{01}/E_n \propto e^{-L/\xi}$.

We have seen that the topological protection of the qubit, due to the separation of the MBS, also holds for the virtual processes in the regime of nearly adiabatic manipulation. 
Therefore, in order to implement a phase gate according to \equref{Translation_of:PhaseOfRho}, one has to either bring the Majorana modes close together such that $L$ is of order $\xi$ or use sufficiently high velocities, where the perturbative approach in $\mathcal{M}$ breaks down. In this paper we study the latter option.

\subsection{Parallel translation of two MBS}
\label{Translation}
From \secref{Limitations} we know that the coherence and the fermion parity of the topological qubit are most stable against large braiding velocities, if the system length $L$ is held constant. For this reason let us now focus on the case $\dot{x}_1(t)=\dot{x}_2(t)=:v(t)$, which, in addition, provides an alternative way of treating non-adiabatic effects.

\subsubsection{Next adiabatic iteration}
To see this, let us introduce the spatial displacement operator
\begin{equation}
 \mathcal{W}(t) = \mathbbm{1}\, e^{-i p \int_{t_0}^t\diff t' v(t')}, \label{Translation_of:TranslationOperator}
\end{equation} 
where $\mathbbm{1}$ denotes the identity matrix both in spin and particle hole space. This enables us to write
\begin{equation}
 h(t) = \mathcal{W}(t) h(t_0) \mathcal{W}^\dag(t)
\end{equation}  
and, consequently, the instantaneous eigenstates satisfy $\ket{\phi_n(t)}=\mathcal{W}(t)\ket{\phi_n(t_0)}$.
Therefore, the corresponding ``moving frame'' BdG Hamiltonian is simply given by
\begin{align}
 h'(v(t)) 	&= h(t_0) - i\,\mathcal{W}^\dag(t)\dot{\mathcal{W}}(t)  \\
		&= h(t_0) - \mathbbm{1}\,v(t)\,p, \label{Translation_of:InstandBdGH2} 
\end{align}
which determines the entire dynamics of the system by means of \equref{General_expressions:TimeEvolutionAdiabaticBasis}. Note that we have found an explicit form of $h'$ without having to calculate entries of the coupling matrix $\mathcal{M}$ explicitly.

Instead of applying perturbation theory in the second term in \equref{Translation_of:InstandBdGH2}, \ie in the velocity $v$, let us follow an approach which essentially goes back to \refcite{Berry3} and diagonalize $h'(v)$ for every $v$,
\begin{equation}
 h'(v)\ket{\phi_n'(v)} = E_n'(v) \ket{\phi_n'(v)}.
\end{equation} 
Since $v$ is only a one-dimensional parameter, we can use again the parallel transport condition, $\braket{\phi_n'|\partial_v\phi_n'} = 0$, to remove the phase ambiguity between the (single-valued) superadiabatic eigenfunctions $\ket{\phi_n'(v)}$ at different values of $v$. Similar to \equref{General_expressions:MovingFrameHamiltonian}, the Hamiltonian governing the time-evolution in the superadiabatic basis is given by 
\begin{equation}
 h''_{n,m}(t) = E'_n(v(t))\delta_{n,m} - \mathcal{M}'_{n,m}(t), \label{Translation_of:SuperAdiabaticHamiltonian}
\end{equation} 
where the new coupling matrix elements read 
\begin{equation}
 \mathcal{M}'_{n,m}(t)=i\,\dot{v}(t)\braket{\phi'_n(v(t))|\partial_v\phi'_m(v(t))}.
\end{equation} 
For any realistic translation process we have $v(t)\rightarrow 0$ as $|t|\rightarrow \infty$. Under this assumption, it is easily seen that the matrix exponential defined in \equref{General_expressions:TimeEvolutionAdiabaticBasis} obeys for $t_0\rightarrow-\infty$ and $t\rightarrow\infty$
\begin{equation}
 u_{n,m} = \left[\mathcal{T} \exp\left(-i\int_{-\infty}^\infty \diff t' \, h''(t')\right) \right]_{n,m}. \label{Translation_of:TimeEvolutionOperator}
\end{equation} 
In straightforward analogy, the perturbative analysis of \secsref{Qubitquantities}{EffectiveTheory} can now be repeated in the superadiabatic basis. Since $\mathcal{M}'\propto \dot{v}$, this yields results which are perturbative in the acceleration but contain $v$ to arbitrary order making the regime of large velocities accessible.

\subsubsection{Quantum wire at large magnetic fields}
\label{TranslationLargeB}
\begin{figure}[t]
\begin{center}
\includegraphics[width=\linewidth]{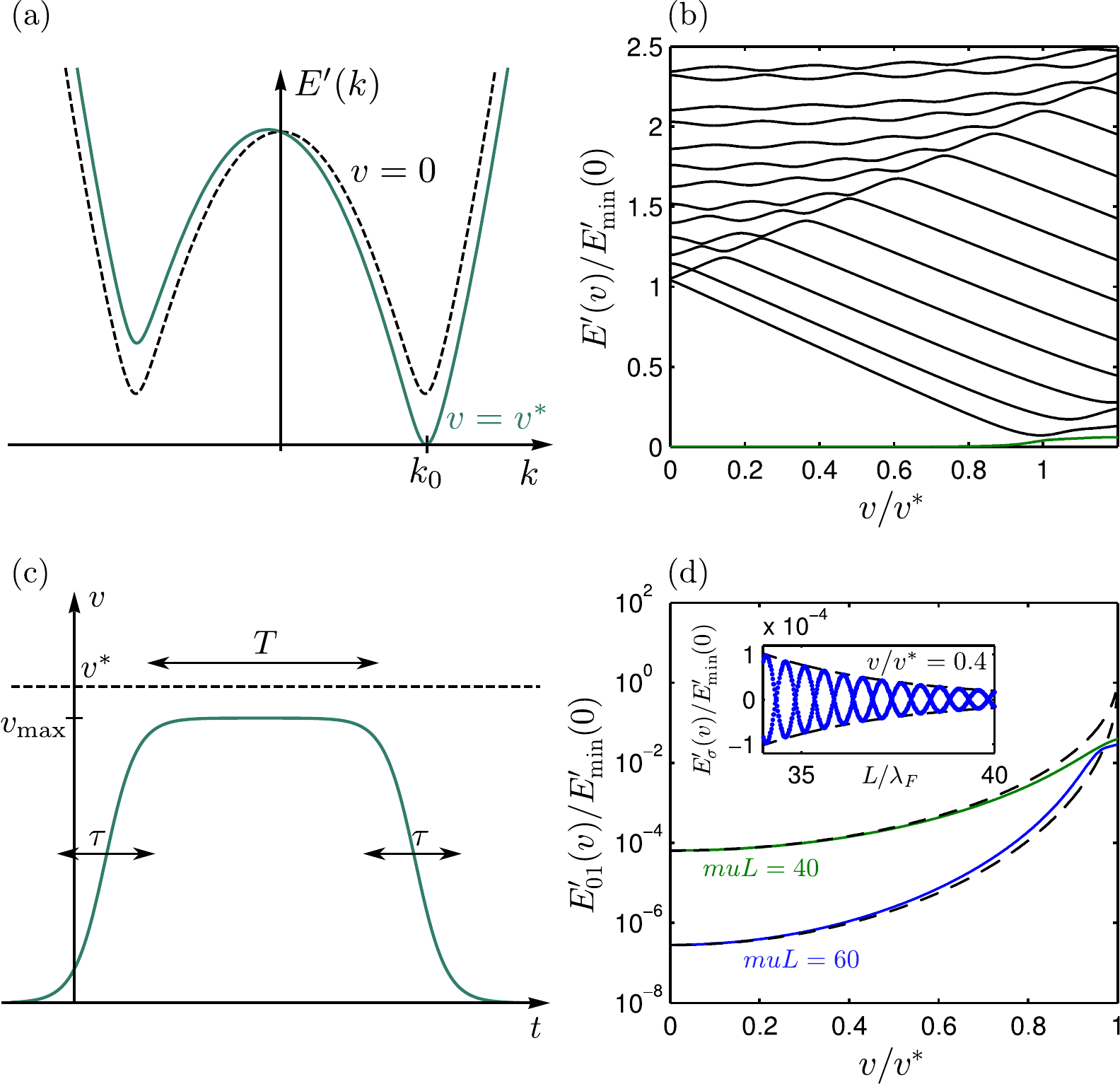}
\caption{(Color online) (a) Excitation spectrum of the ``moving frame'' Kitaev model (\ref{Translation_of:MovingKitaev}) for $v=0$ (black dashed line) and right at the critical velocity $v^*$ (green line) at which the system becomes gapless. In (b) numerical results for the energy splitting (green) and for the energies of the first states above the gap (black) as a function of the velocity $v$ are shown. Here we used the ``moving frame'' Hamiltonian of the wire Hamiltonian (\ref{Majorana_Fermions:WireBdG}) with $B/\Delta=4$, $B/\epsilon_{\text{so}}=25$, $muL=40$ and $\mu_2=-10B$. Part (c) illustrates the suggested velocity profile for realizing a non-adiabatic phase gate. In (d) we show numerical results for the energies of the ground state fermions as a function of the velocity $v$ for two different system lengths and as a function of $L$ for a fixed value of $v$ (inset). Here the same parameters as in (b) have been used. The black dashed lines correspond to the expected scaling behavior in \equref{Translation_of:ExpectedScaling}.}
\label{QWTransPlot}
\end{center}
\end{figure}

Again, let us focus on the regime of strong magnetic fields, $B\gg \epsilon_{\text{so}},\Delta$, where the system can be effectively described by Kitaev's model (\ref{Quantum_wire:ContKitModel2}) and hence the ``moving frame'' Hamiltonian reads
\begin{equation}
 h'_{\text{Kit}}(v)   = \left(\frac{p^2}{2m} - \mu_e(x) \right)\tau_z - v_e p \,\tau_y - \mathbbm{1}\,v\,p. \label{Translation_of:MovingKitaev}
\end{equation}
The resulting excitation spectrum $E'(k)$ is illustrated in \figref{QWTransPlot}(a).
We observe that the presence of the additional term $vp$ in the BdG Hamiltonian tilts the dispersion and drives the system into a gapless phase for
\begin{equation}
v > v^* \approx v_e \approx \frac{u\Delta}{B}. \label{Translation_of:CriticalVelocity}
\end{equation} 
The system is very sensitive to small accelerations when the velocity $v^*$ is reached, since the superadiabatic gap $E'_\text{min}(v)$ closes for $v\rightarrow v^*$. Therefore, we recover the concept of a critical velocity scale also for the entirely translated system. In the present case, however, the qubit will still be protected from decoherence for sufficiently small $\dot{v}$, even if $v$ becomes comparable with $v^*$.

To understand how the superadiabatic MBS behave when the velocity $v$ is increased, let us investigate the four possible wavevectors $q'=\{\pm q'_+,\pm q'_-\}$ of the Majorana wavefunctions. For constant $\mu_e(x)=\mu_1>0$ one finds from \equref{Translation_of:MovingKitaev} 
\begin{equation}
 q'_\sigma(v) \approx mv_e\sqrt{1-\left(v/v_e\right)^2} + \sigma\, i\sqrt{2m\,\mu_1}, \label{Translation_of:MovingWVApprox}
\end{equation} 
for vanishing tunnel splitting $E'_{01}(v)=0$, \ie assuming an infinitely long topological domain. Since $\Re{q'_\sigma(v)}$ decreases with $v$, the MBS wavefunctions broaden when the system is accelerated. The tunnel splitting becomes significant above a certain velocity $v<v^*$ for any finite length $L$ of the topological domain because  $\Re{q'_\sigma(v)}$ approaches $0$ as $v\rightarrow v^*$. We expect that the associated ground state eigenfunctions $\ket{\phi'_{01}}$ and $\ket{\phi'_{\cbar{01}}}$ finally merge into the continuum for $v\gtrsim v^*$. 

For a more quantitative picture, the ``moving frame'' Hamiltonian $h'(v)$ in \equref{Translation_of:InstandBdGH2} is diagonalized numerically using the full four component BdG Hamiltonian (\ref{Majorana_Fermions:WireBdG}) of the nanowire. We again take a piecewise constant chemical potential to describe the domain walls between the topological ($\mu=0$) and trivial ($\mu=\mu_2<-\sqrt{B^2-\Delta^2}$) phases illustrated in \figref{Dispersion3DPlot}(a).

\figref{QWTransPlot}(b) shows the resulting energies of the first excited states above the gap and the tunnel splitting $E'_{01}$ as a function of the velocity $v$. As expected, the energies of the excited states are reduced with increasing $v$ (with a slope approximately given by $k_F$ in the linear regime) and the fermionic ground state solutions of $h'(0)$ smoothly connect to the ordinary states of the quasi-continuum at $v\approx v^*$.  

For a concrete implementation of a phase gate let us investigate the protocol illustrated in \figref{QWTransPlot}(c), which is characterized by the two time scales $\tau$ and $T$ representing the acceleration time and the time during which the maximum velocity $v_{\text{max}}<v^*$ is held constant.

In analogy to \equref{Translation_of:PhaseOfRho}, the phase $\varphi$ accumulated during the processes can be written as $\varphi = \varphi_1 + \varphi_2$, where
\begin{equation}
 \varphi_1 = \int_{-\infty}^\infty\mathrm{d}t_1\, E'_{01}(v(t_1)) \label{Translation_of:DynamicalPartOfPhase}
\end{equation} 
is the dynamical phase in the superadiabatic basis and $\varphi_2$ the acceleration correction. Applying the approximative projection procedure presented in \secref{EffectiveTheory} to the superadiabatic Hamiltonian (\ref{Translation_of:SuperAdiabaticHamiltonian}), one finds 
\begin{equation}
 \varphi_2 \approx -\sum_{n\neq 0}\int_{-\infty}^\infty \diff t \,\,\dot{v}^2\,\frac{\left|\braket{\phi_{01}'(v)|\partial_v\phi'_n(v)}\right|^2}{E'_n(v)-E'_{01}(v)} \label{Translation_of:PhaseCorrection}
\end{equation} 
valid for small $\dot{v}$. 

To begin with the dynamical phase $\varphi_1$, numerical results for the tunnel splitting $E'_{01}(v)$ as a function of the velocity are shown in \figref{QWTransPlot}(d) for two different values of the system length $L$. As expected from the analysis of the virtual energy correction (\ref{Translation_of:dEThird}) in the adiabatic basis, large velocities comparable with $v^*$ are to be applied to generate a significant increase in $E_{01}'(v)$. Since the gate operation time has to be much shorter than the time scale $1/E'_{01}(0)$ on which the phase of $\rho^Q_{01}$ is modified due to the tunnel splitting at $v=0$, we require $E_{01}'(v_{\text{max}})/E_{01}'(0)\gg 1$. 
Focusing on $muL=60$, the numerical data in \figref{QWTransPlot}(d) implies that one has to choose $v_{\text{max}}/v^*=0.8$ such that $E_{01}'(v_{\text{max}})/E_{01}'(0)$ is of order $10^2-10^3$.

At least for sufficiently small $E_{01}'(v)$, the envelope function of $E_{01}'$ is expected to scale as $\exp{(-\Re{q'_\sigma(v)}L})$, where $q'_\sigma(v)$ is given by \equref{Translation_of:MovingWVApprox}.
Fitting the prefactor to the numerical data yields good agreement for the scaling behavior of $E_{01}'$ as a function of both $v$ and $L$ as can be seen from the black dashed lines in \figref{QWTransPlot}(d).
For that reason,
\begin{equation}
 \frac{E_{01}'(v_{\text{max}})}{E_{01}'(0)} \propto e^{\left(\Re{q'_\sigma(0)} - \Re{q'_\sigma(v_{\text{max}})} \right)L} \label{Translation_of:ExpectedScaling}
\end{equation}      
is to be expected and hence the enhancement of the energy splitting can be easily enlarged by increasing the system length $L$.

To avoid decoherence and parity errors, $\tau$ has to be chosen much larger than the time scale $\tau_c=1/E_{\text{min}}'(v_{\text{max}})$. Assuming that $v_{\text{max}}/v^*=0.8$, we find $\tau_{c}\approx 5/E'_{\text{min}}(0)$ from \figref{QWTransPlot}(b). In order to investigate the relevance of the contribution of $\varphi_2$ we have calculated the scalar products and the summation in \equref{Translation_of:PhaseCorrection} numerically. Again referring to $v_{\text{max}}/v^*=0.8$ and $muL=60$, one finds $\varphi_2/(2\pi) < 10^{-2}$ for $\tau \gtrsim 10^2\tau_c$. Consequently, with these parameters not only the coherence and the parity of the qubit are expected to be unaffected, but also $\varphi \approx E_{01}'(v_{\text{max}})T$ with corrections due to the acceleration which are smaller by a factor of order $10^{-2}$. In principle, however, one could also take the acceleration corrections into account by properly calibrating the phase gate.

The inset in \figref{QWTransPlot}(d) shows the energies $E'_{\sigma}(v)$, $\sigma=01,\overline{01}$, for $v=0.4\,v^*$ as a function of $L$. We observe that the superadiabatic tunnel splitting is sinusoidal in $L$ with periodicity given by the Fermi wavelength $\lambda_F$ as is well-known for the adiabatic tunnel splitting.\cite{PhaseGate,ChengMeng} 
This means that $L$ has to be stabilized with an accuracy on the length scale $\lambda_F$ for a controllable manipulation of the phase of $\rho^Q_{01}$. 
Note that this problem can be overcome in the scheme based on the direct overlap of the MBS wavefunctions by using their monotonically decaying behavior in the non-topological domains of the wire. \cite{PhaseGate} Unfortunately, this solution of the problem is not readily transferred to our proposal. Nevertheless, we believe that the notion of a non-adiabatic phase gate may not only be of theoretical interest because it represents an alternative that might be helpful for making potential quantum computation schemes to work more efficiently. Furthermore, the approach presented here may be seen as a starting point for the search of more sophisticated procedures of constructing gate operations based on non-adiabatic effects. In the following subsection we present a first example of a proposal that is easier accessible experimentally.

\subsection{Manipulations by supercurrent}
\label{ComparisonwSC}
The analysis presented above has shown that a phase gate which does not require bringing the MBS close to each other is in principle possible by taking advantage of non-adiabatic effects.
Although the error threshold is very large ($14\%$) due to a correction scheme known as ``magic-state distillation'',\cite{BravyiKitaev,Bravyi} the experimental implementation may be complicated  since high velocities of order $v^*$ are required, while $L$ has to be fixed with high accuracy ($\delta L \ll \lambda_F$). 
Therefore, let us investigate a related setup, first studied in \refcite{Romito}, where instead of moving the MBS relative to the $s$-wave superconductor, a supercurrent $J(t)$ is driven through the proximity inducing superconductor along the nanowire.

The presence of the current induces a gradient $\partial_x\varphi(x,t)$ in the phase of the superconducting order parameter $\Delta$ in \equref{Majorana_Fermions:WireBdG} according to $J(t)\propto \partial_x\varphi(x,t)$. Focusing, on the case where $J(t)$ is spatially uniform and applying a suitable gauge transformation,\cite{Romito} the Hamiltonian can be written as   
\begin{align}
\begin{split}
  &\widetilde{h}_{\text{NW}}(\partial_x\varphi) = \Biggl(\frac{1}{2m}\left(p^2+\left(\frac{\partial_x\varphi}{2}\right)^2\right)-\mu  \\
&+u \biggl(p -\tau_z\frac{\partial_x\varphi}{2} \biggr) \sigma_y\Biggr)\tau_z  + B \sigma_z - \Delta \tau_x  - \frac{\partial_x\varphi}{2m}p. \label{Translation_of:WireHamiltonianWithCurrent2}\end{split}
\end{align} 
For a derivation of the resulting phase diagram for time-independent currents we refer the reader to \refcite{Romito}. Here we will only discuss the applicability of this system for implementing a phase gate.

\subsubsection{Limit of large magnetic fields}
As before, we first analyze the regime of strong magnetic fields. It is instructive to compare \equref{Translation_of:WireHamiltonianWithCurrent2} with the ``moving frame'' Hamiltonian
\begin{equation}
 h'_{\text{NW}}(v) = \left(\frac{p^2}{2m} - \mu + u p \,\sigma_y\right)\tau_z + B \sigma_z - \Delta \tau_x - v\,p \label{Translation_of:MovingFrameHamFourComp}
\end{equation} 
for the spatially displaced topological domain. We observe that the last terms in \equsref{Translation_of:WireHamiltonianWithCurrent2}{Translation_of:MovingFrameHamFourComp} are identical upon identifying $v=\partial_x\varphi/(2m)$. For low energies $E\ll B$ and $v<v^*$, the momenta of the excited states of $h'_{\text{NW}}(v)$ are of order of the Fermi wavevector $k_F\approx\sqrt{2mB}$ (see \figref{QWTransPlot}(a)). Since $|q'_\sigma(v)|\gtrsim k_F$, the Hamiltonians in \equsref{Translation_of:WireHamiltonianWithCurrent2}{Translation_of:MovingFrameHamFourComp} are identical for low energies as long as $\partial_x\varphi\ll k_F$. According to the identification $v=\partial_x\varphi/(2m)$ and using \equref{Translation_of:CriticalVelocity}, the critical phase gradient is given by 
\begin{equation}
 (\partial_x\varphi)^*_2 = 2mv^* \approx \frac{2mu\Delta}{B},
\end{equation}
which indeed satisfies $(\partial_x\varphi)^*_2\ll k_F$. 
 
For this reason, $\widetilde{h}_{\text{NW}}(\partial_x\varphi)$ and $h'_{\text{NW}}(v)$ with $v=\partial_x\varphi/(2m)$ exhibit the identical low energy behavior for the entire topological regime.
Therefore, we know that the system undergoes a transition at $(\partial_x\varphi)^*_2$ from a topological phase into a gapless phase in accordance with the results of \refcite{Romito}. 
Moreover, the numerical data shown in \figref{QWTransPlot} and hence all estimates in \secref{Translation} also hold for this setup upon replacing $v/v^*$ by $\partial_x\varphi/(\partial_x\varphi)^*_2$. As a consequence, supercurrents can be used similar to the translation of the topological domain for realizing a phase gate without bringing MBS closer together. 

On top of that, we believe that driving supercurrents along the nanowire may be easier to implement experimentally, since no local gate tuning is required. Furthermore, the stabilization of the length of the topological domain is expected to be less problematic, if the position of both MBS is fixed.

\subsubsection{Current induced topological phase}
\begin{figure}[t]
\begin{center}
\includegraphics[width=\linewidth]{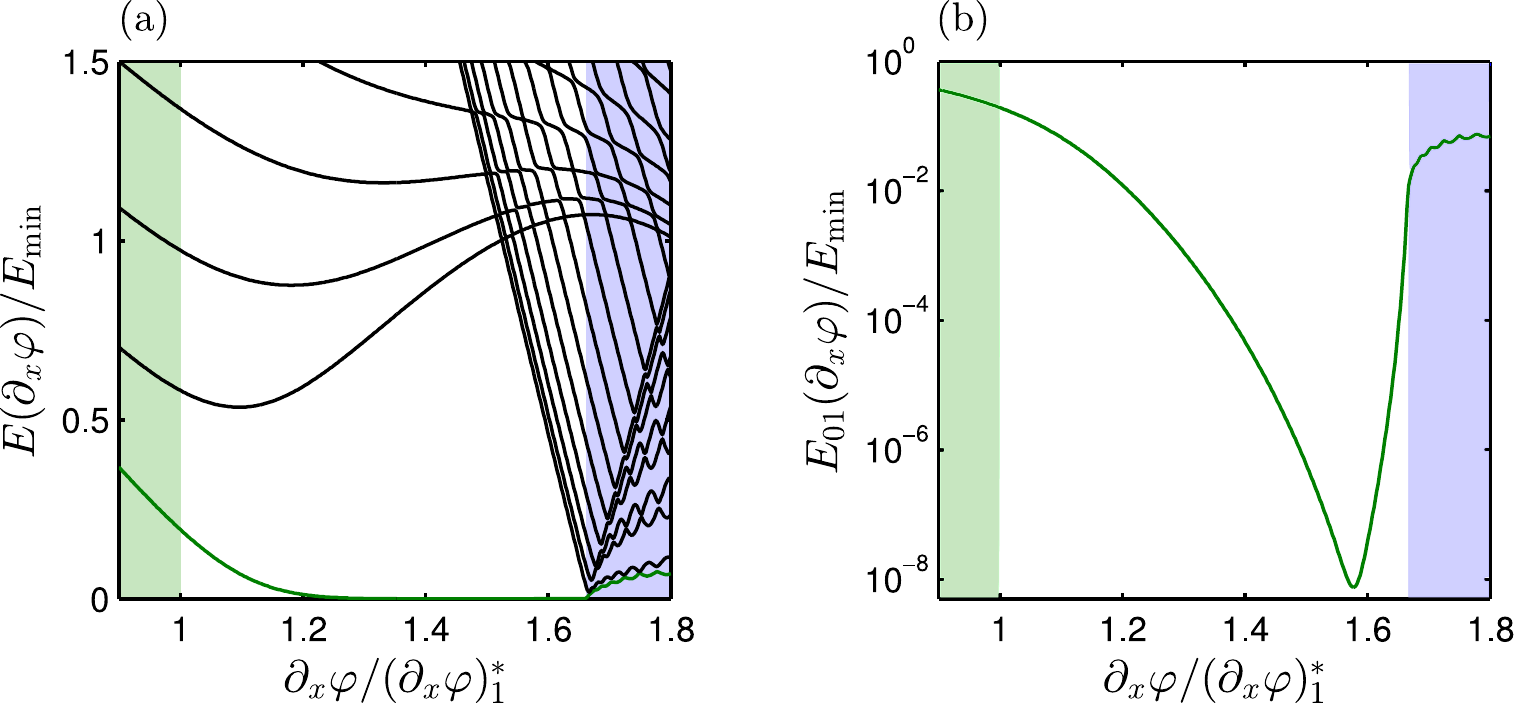}
\caption{(Color online) (a) Excitation energies of \equref{Translation_of:WireHamiltonianWithCurrent2} for different values of the phase gradient $\partial_x\varphi$ obtained numerically using $B/\Delta=0.9$, $\epsilon_{\text{so}}/\Delta=0.5$, $muL=80$ and taking $\mu=0$ in the topological segment of the wire. The trivial and gapless phases are highlighted in green and blue, respectively. $E_{\text{min}}$ is defined as the maximum gap in the topological phase. In (b) the tunnel splitting is shown again on a logarithmic scale.}
\label{CurrentInducedPlot}
\end{center}
\end{figure}

By applying a supercurrent along the nanowire it is even possible to stabilize a topological phase when $B<\Delta$, for which the system would reside in the trivial phase, if no current was present. Clearly this cannot be achieved by simply translating the system, since the additional term $vp$ in \equref{Translation_of:InstandBdGH2} leaves the topological gap at $k=0$ unaffected. 

In \figref{CurrentInducedPlot}(a) numerical results for the excitation energies of the BdG Hamiltonian (\ref{Translation_of:WireHamiltonianWithCurrent2}) with $B/\Delta=0.9$ and $\epsilon_{\text{so}}/\Delta=0.5$ are shown as a function of the gradient $\partial_x\varphi$. 
We know from \refcite{Romito} that for these parameters one can realize a trivial ($\partial_x\varphi < (\partial_x\varphi)^*_1$), a topological ($(\partial_x\varphi)^*_1<\partial_x\varphi < (\partial_x\varphi)^*_2$) and a gapless phase ($\partial_x\varphi > (\partial_x\varphi)^*_2$) by tuning the strength of the phase gradient. As expected, the quasi-zero energy modes of the topological domain merge into the continuum for $\partial_x\varphi\lesssim (\partial_x\varphi)^*_1$ and $\partial_x\varphi\gtrsim (\partial_x\varphi)^*_2$. On the one hand, the transition to the gapless phase at $(\partial_x\varphi)^*_2$ is similar to the behavior found in \figref{QWTransPlot}(b) for the system displaced with velocity $v\lesssim v^*$. The gap is strongly reduced for $\partial_x\varphi\rightarrow (\partial_x\varphi)^*_2$ compared to its maximum value $E_{\text{min}}$ at $\partial_x\varphi\approx 1.5(\partial_x\varphi)^*_1$. On the other hand, when $\partial_x\varphi$ approaches  $(\partial_x\varphi)^*_1$ from above, the gap is reduced only by a factor of $2$ making the qubit much less sensitive to parity errors and decoherence effects. Clearly, this is a finite size effect and reflects the fact that the density of states of the non-topological phase is much smaller than that of the gapless phase. However, as far as the topological protection of the qubit is concerned, the system length only has to be sufficiently large compared to the localization length of the MBS such that the tunnel splitting $E_{01}$ is negligible for the purpose of information storage. From \figref{CurrentInducedPlot}(b) we can see that $E_{01}$ assumes a minimum value of $10^{-8}E_{\text{min}}$ slightly below $\partial_x\varphi=1.6(\partial_x\varphi)^*_1$. By tuning the supercurrent to this minimum, the topological memory can be stored on a time-scale which is $7-8$ orders of magnitude larger than the scale $1/E((\partial_x\varphi)^*_1)\approx 4/E_{\text{min}}$ associated with adiabaticity during a phase gate operation based on approaching the boundary $(\partial_x\varphi)^*_1$ to the non-topological phase.

\section{Conclusions}
In this work we have studied the non-adiabatic dynamics of a topological qubit due to the presence of a quasi-continuum of degrees of freedom above the gap of the system. 
To characterize the qubit, the off-diagonal component $\rho^Q_{01}$ of its reduced density matrix as well as the parity $\hat{\mathcal{P}}_{01}$ of the fermion associated with $\hat{\gamma}_1$ and $\hat{\gamma}_2$ have been considered. We discussed both aspects of non-adiabatic effects, the limitations they impose on topological quantum computation and the opportunities they provide for creating additional gate operations. 

First, a general class $D$ topological superconductor coupled to a time-dependent classical field has been investigated and perturbative expressions for $\braket{\hat{\mathcal{P}}_{01}}$ and $\rho^Q_{01}$ have been derived. Our results show that correlation effects between remote Majorana modes, mediated by the extended states of the continuum, can lead to the accumulation of a phase by the qubit, even if the overlap of their wavefunctions is small.  
The derivation of an effective ground state Hamiltonian for the regime of nearly adiabatic manipulation makes it clear that this phase can be seen as the result of the non-adiabatic renormalization of the tunnel splitting.

In the remainder of the paper, these generic results have been applied to the spin-orbit coupled nanowire in proximity to an $s$-wave superconductor assuming that the MBS are transported by varying the chemical potential profile. Focusing on the limit of large magnetic fields, we used Kitaev's model to calculate the relevant coupling matrix elements which turned out to be oscillatory as a function of the length $L$ of the topological domain (with periodicity $2\lambda_F$). This implies a critical velocity scale $v_c\approx 2E_{\text{min}}/k_F$ for moving one edge of the topological segment of the wire. Our analysis reveals that the resulting lower bound on the braiding time is expected to be of the same order as the upper bound imposed by single-electron tunneling.\cite{LossPoisoning} By taking a prototypal analytic trajectory $L(t)$, we have shown that the non-adiabatic corrections are exponentially suppressed ($\propto\exp(-2E_{\text{min}}\tau)$) for small velocities $\dot{L}\ll v_c$.
In the regime of strong spin-orbit coupling we found a similar oscillatory behavior of the coupling matrix elements and the resulting lower bound on the braiding time turned out to be of the same order as in the case of strong magnetic fields. 

We then investigated, in detail, the possibility of constructing a non-adiabatic phase gate. Our findings show that a perturbative treatment of the braiding velocities $\dot{x}_j$ of the MBS only leads to non-adiabatic corrections to the tunnel splitting that are exponentially small ($\exp(-L/\xi)$) in the distance $L$ between the Majorana modes. 
In order to investigate larger braiding velocities, we assumed that the entire topological domain is translated making a perturbation approach in the acceleration possible. We found a critical velocity $v^*$, of the same order as $v_c$ above, where the effective ``moving frame'' Hamiltonian becomes gapless. A trajectory is presented where the relative error of the accumulated phase due to the acceleration parts of the protocol and decoherence effects as well as parity errors are expected to be negligible.
However, engineering a non-adiabatic phase gate by translation of the entire topological domain may be difficult in practice, since the distance between the MBS has to be well-stabilized on the length scale $\lambda_F$. 
This is why we analyzed a different setup,\cite{Romito} where the position of the MBS is fixed and, instead, a supercurrent is applied along the nanowire.
In the regime of strong magnetic fields, the two systems are essentially identical and hence supercurrents can be used similar to the translation of the topological domain for implementing a phase gate. On top of that, supercurrents reveal additional ways for accumulating a phase. For example, in the parameter regime where a topological phase can be stabilized by a current even if $B<\Delta$, we found that finite size effects can be exploited for efficiently protecting the qubit from decoherence.
This shows already that there are various possibilities to improve the most straightforward approach of just translating MBS in order to construct non-adiabatic gate operations.
In fact, we believe that this paper might pave the way for the development of more elaborate schemes for the implementation of non-adiabatic phase gates or even of non-trivial two-qubit gates.

\begin{acknowledgments}
We gratefully thank Ivar Martin, Yuriy Makhlin, Panagiotis Kotetes and Bhilahari Jeevanesan for various inspiring and useful discussions.
AS is grateful to the Weizmann Institute of Schience (Weston Visiting Professorship) and the BMBF Project RUS 10/053 ''Topologische Materialien f\"ur Nanoelektronik''.
\end{acknowledgments}

\textit{Note added.} -- \,During the finial steps of the preparation of this manuscript we became aware of \refcite{vonOppenSim} which shows considerable overlap with our work.
\appendix
\section{Geometric matrix elements}
\label{Ap:MatrixElements}
In this appendix we present the analytical result for the geometric coupling matrix elements $b_{\gamma_2,n}=\braket{\phi_{\gamma_2}|\partial_{x_2} \phi_n}$ using the Kitaev model (\ref{Quantum_wire:ContKitModel2}) and an infinite potential well.

We introduce the pseudo parity operator $\Pi = \tau_z \mathcal{P}$, where $\mathcal{P}$ denotes the spatial inversion with respect to the center of the topological domain, and take advantage of $\Pi^2=\mathbbm{1}$ as well as the fact that $\Pi$ commutes with the Hamiltonian. Within each of the two eigenspaces of $\Pi$ associated with eigenvalues $\lambda=\pm 1$, there is a sequence of continuum states with non-degenerate energies $E_s^\lambda$, $s\in\mathbbm{N}^+$.
Let us chose the sequence $E_s^\lambda$ to be monotonically increasing and define the auxiliary quantum number $\sigma=(-1)^s\lambda$, which turns out to be central in the calculation of $b_{\gamma_2,n}$.
Since it is not possible to find an analytical expression for the discrete energies $E_s^\lambda$, we treat the energy of the continuum states as a continuous variable $E$ and calculate a smooth interpolating function $b_{\gamma_2}(E)$.
In the limit $L \gg \xi,\lambda_F$, we obtain
\begin{widetext}
\begin{align}
\begin{split}
 b_{\gamma_2}(E,L) 	&\approx \frac{\sqrt{2v_e \mu_1}}{\sqrt{L\left(1+R^2(E,L)\right)}\,E} \Biggl[k_-(E)\, \sin\left(\left(k_F - \delta k(E,L)\right)\frac{L}{2}+\varphi_-(E)\right) \\
			&\qquad+ R(E,L)\,k_+(E)\, \sin\left(\left(k_F + \delta k(E,L)\right)\frac{L}{2}+\varphi_+(E)\right) \Biggr], \label{Quantum_wire:MatrixElementsResults} \end{split}
\end{align} 
where
\begin{equation}
 \varphi_\pm(E) := \frac{\theta(k_\pm(E))}{2} +\arctan\left(\frac{4m^2v_e^2+\Omega^2-k_\pm^2(E)}{4mv_ek_\pm(E)} \right)
\end{equation}
and 
\begin{align}
\begin{split}
\delta k(E,L) &:= \frac{\pi}{L}\,\delta_{\sigma,-1} + \frac{2\,\text{sign}(\sigma)}{L}\arctan\Biggl(\frac{1}{2(1-g)\tan\hspace{-0.2em}\left(k_F\frac{L}{2}\right)}\Biggl[ (1+g)\left(1+\tan^2\hspace{-0.2em}\left(k_F\frac{L}{2}\right)\right) \\
		&-\sqrt{(1+g)^2\left(1+\tan^2\hspace{-0.2em}\left(k_F\frac{L}{2}\right)\right)^2-4(1-g)^2\tan^2\hspace{-0.2em}\left(k_F\frac{L}{2}\right)} \Biggr]  \Biggr).\label{Quantum_wire:KappaDefinition}\end{split}
\end{align}
\end{widetext}
In \equstref{Quantum_wire:MatrixElementsResults}{Quantum_wire:KappaDefinition} the following conventions are being used: The angle $\theta(k)\in[0,\pi)$ is defined via
\begin{equation}
 \cos(\theta(k)) = \frac{\mu_1 - \frac{k^2}{2m}}{\epsilon(k)},
\end{equation} 
where 
\begin{equation}
 \epsilon(k) = \sqrt{\left(\frac{k^2}{2m}-\mu_1\right)^2 + v_e^2 k^2} \label{Quantum_wire:Dispersion}
\end{equation}
is the spectrum of (\ref{Quantum_wire:ContKitModel2}). The Fermi wavevector is given by
\begin{equation}
 k_F = \sqrt{2m(\mu_1-m v_e^2)}
\end{equation}
and coincides with the real part $\Omega$ of the wavevectors of the MBS in the physically relevant limit of large magnetic fields ($\mu_1\gg mv_e^2$). Furthermore,
\begin{align}
k_{\pm}(E)=\sqrt{k_F^2 \pm 2m\sqrt{E^2-E_{\text{min}}^2}}
\end{align}
are the absolute values of the four possible momenta $\{\pm k_+,\pm k_-\}$ of a continuum state with energy $E<\mu_1$.
In addition, we defined
\begin{equation}
 R = -\frac{\cos\left(\frac{\theta(k_F-\delta k)}{2}\right)}{\cos\left(\frac{\theta(k_F+\delta k)}{2}\right)}\, \frac{\sin\left((k_F-\delta k)\frac{L}{2}\right)}{\sin\left((k_F+\delta k)\frac{L}{2}\right)}
\end{equation}
and 
\begin{equation}
  g(E) =  \frac{\tan(\theta(k_-(E))/2)}{\tan(\theta(k_+(E))/2)}. \label{Quantum_wire:gDefinition}
\end{equation} 
In \figref{Dispersion3DPlot}(e) this result is shown for $\sigma=+1$. In the second case, $\sigma=-1$, the geometric matrix elements are identical except for a shift by $\lambda_F/2$ in their functional dependence on $L$. 

Exploiting again the pseudo inversion symmetry of the problem, one can easily show that the coupling matrix elements involving the left MBS $\gamma_1$ are simply given by $b_{\gamma_1,n}=\pm i \lambda b_{\gamma_2,n}$, where the constant sign $\pm$ depends on how the relative sign between the wavefunctions $\phi_{\gamma_1}$ and $\phi_{\gamma_2}$ is chosen.

\bibliography{paper.bib}
\end{document}